\documentclass[%
 aip,
 jcp,%
 amsmath,amssymb,
reprint,%
]{revtex4-1}

\usepackage{epsfig}
\usepackage{color}
\usepackage{graphics}
\usepackage{graphicx}
\usepackage{amsmath}
\usepackage[english]{babel}
\begin{document}

\hyphenation{ALPGEN}
\hyphenation{EVTGEN}
\hyphenation{PYTHIA}

\title{The effect of soft repulsive interactions on the diffusion of particles in quasi-one-dimensional channels: A hopping time approach.}
\author{Sheida Ahmadi}
\affiliation{Department of Chemistry, University of Saskatchewan, Saskatoon, Saskatchewan S7N 5C9, Canada}

\author{Marina Schmidt}
\affiliation{Department of Computer Science, University of Saskatchewan, Saskatoon, Saskatchewan S7N 5C9, Canada}

\author{Raymond J. Spiteri}
\affiliation{Department of Computer Science, University of Saskatchewan, Saskatoon, Saskatchewan S7N 5C9, Canada}

\author{Richard K. Bowles}
\email{richard.bowles@usask.ca}
\affiliation{Department of Chemistry, University of Saskatchewan, Saskatoon, Saskatchewan S7N 5C9, Canada}

\begin{abstract}
Fluids confined to quasi-one-dimensional channels exhibit a dynamic crossover from single file diffusion to normal diffusion as the channel becomes wide enough for particles to hop past each other. In the crossover regime, where hopping events are rare, the diffusion coefficient in the long time limit can be related to a hopping time that measures the average time it takes a particle to escape the local cage formed by its neighbours. In this work, we show that a transition state theory that calculates the free energy barrier for two particles attempting to pass each other in the small system isobaric ensemble is able to quantitatively predict the hopping time in a system of two-dimensional soft repulsive discs [$U(r_{ij})=(\sigma/r_{ij})^{\alpha}$] confined to a hard walled channel over a range of channel radii and degrees of particle softness measured in terms of $1/\alpha$. The free energy barrier exhibits a maximum at intermediate values of $\alpha$ that moves to smaller values of $1/\alpha$ (harder particles) as the channel becomes narrower. However, the presence of the maximum is only observed in the hopping times for wide channels because the interaction potential dependence of the kinetic prefactor plays an increasingly important role for narrower channels. We also begin to explore how our transition state theory approach can be used to optimize and control dynamics in confined quasi-one-dimensional fluids.
\end{abstract}

\maketitle

\section{Introduction}
\label{sec:intro}
The movement of fluids through narrow quasi-one-dimensional channels appears in a variety of natural and engineered systems~\cite{Karger_Transport_2015,Gubbins_The_2010} such as zeolites,~\cite{Kukla_NMR_1996,Karger_Single_1992,Kumar_Crossover_2014} carbon nanotubes,~\cite{Das_Single_2010,Valiullin_Comment_2010, Chen_Transition_2010} metallic organic frameworks,~\cite{Salles_Molecular_2011,Jobic_Observation_2016} confined colloids,~\cite{Wei_Single_2000,Lutz_Single_2004} and in the transport of water and ions through membranes.~\cite{Hodgkins_The_1955,Finkelstein_Water_1987} It also plays a fundamental role in the functioning of nano- and micro- fluidic devices~\cite{Siems_Non_2012,Locatelli_Single_2016,Yang_Single_2010} and provides a basis for separating mixtures.~\cite{Adhangale_Exploiting_2003,Ball_Normal_2009,Wanasundara_Single_2012}  When the particles are subjected to stochastic forces~\cite{Levitt_Dynamics_1973,Mon_Molecular_2003} or a Brownian background~\cite{Percus_Anomalous_1974} and the channel is narrow enough to prevent passing, the system exhibits a form of sub-diffusion known as single file diffusion (SFD), where the mean squared displacement (MSD) increases as the square root of time in long time limit. The MSD can then be described by an Einstein-like relation,
\begin{equation}
\langle \Delta x_{t}^{2} \rangle=2F_{x}t^{1/2}\mbox{,}
\label{eq:SFD}
\end{equation}
where $F_x$ is the mobility factor.~\cite{Hahn_Deviations_1998,Lin_From_2005}  As the channel becomes wider, the particles are able to pass each other, and the system exhibits a dynamical transition or crossover from SFD to normal diffusion,~\cite{Kumar_Crossover_2014,Chen_Transition_2010,Lucena_Transition_2012,Herrera-Velarde_One_2016,Ooshida_Collective_2018} where the MSD increases linearly in time in the long time limit.  However, in the crossover regime, hopping events are rare because particles must overcome an entropic barrier caused by a restriction in configuration space as they attempt to pass. If the particles remain caged between their neighbours long enough to perform SFD before escaping, the MSD follows Eq.~(\ref{eq:SFD}) at intermediate times before crossing over to normal diffusion at long times.~\cite{Wanasundara_A_2014} 

Percus and Mon~\cite{Mon_Self_2002} developed a simple theory that connects the intermediate time behaviour to the long time normal diffusion coefficient, $D_x$, as,
\begin{equation}
D_x\propto\frac{1}{\tau_{hop}^{1/2}}\mbox{,}\\
\label{eq:Dx}
\end{equation}
where $\tau_{hop}$ is a phenomenological hopping time that measures the average time it takes for a particle to escape its cage.  This hopping time approach to understanding diffusion in confined environments is attractive because all the factors that influence particle mobility, such as density, temperature, and particle--particle and particle--wall interactions, are captured in a single parameter that can be measured directly in simulation. A variety of theoretical and computational methods can be used to study the hopping time. For example, projection operator techniques have been used to map the diffusion of the system into a modified one-dimensional Fick--Jacobs equation~\cite{Kalinay_Projection_2005,Kalinay_Calculation_2007,Kalinay_Two_2008,Mon_Brownian_2008,Mon_Comment_2008,Mon_Brownian_2009} to take advantage of the quasi-one-dimensional nature of the channels. However, the activated nature of the hopping process also suggests $\tau_{hop}$ can be obtained theoretically using barrier crossing methods, and a simple transition state theory has been shown to provide the correct scaling exponent for the hopping time as a function of channel radius~\cite{Bowles_Calculating_2004} for two-dimensional hard discs.  It was recently shown that a transition state theory (TST) qualitatively predicts the hopping time for the same system,~\cite{Ahmadi_Diffusion_2017} where the height of the free energy barrier for two particles attempting to pass was obtained using the small system isobaric-isothermal ensemble.~\cite{Corti_Deriving_1998,Corti_Isothermal_2001}

The goals of the current paper are to show that TST provides quantitative measures of the hopping time for particles confined to quasi-one-dimensional channels when both the barrier and prefactor are calculated and to demonstrate how the hopping time approach can be used to study the role of particle--particle interaction on diffusion in confined fluids. To achieve these goals, we compare the hopping times obtained using TST with those measured directly from the simulation of large number of particles for a system of two-dimensional soft repulsive discs confined to a hard wall channel for a range of channel radii. We also examine the effect of particle softness on the hopping time, and hence diffusion, by varying the magnitude of the exponent associated with the repulsive particle-particle potential. Intuitively, we might expect the hard particle system to exhibit the longest hopping times because they restrict the configuration space of the transition state to the greatest degree. Making the potential softer should make it easier for the particles to reach the transition state. Our results show that the hopping barrier goes through a maximum as the particles become softer because of a competition between the energetic and entropic contributions to the free energy of the transition state. However, the hopping times themselves only appear to exhibit the maximum as a function of particle softness for wider channels because the potential dependence of the kinetic prefactor in the TST theory becomes increasingly important for the narrower channels. We also begin to explore how our TST method can be used to optimize the dynamic properties of single file systems.

The remainder of the paper is organized as follows. Section~\ref{sec:methods} outlines the model system and the methods, including outlining the general TST approach to hopping times, the hopping free energy barrier calculations, the kinetic prefactor calculations, the hopping times measurements in large systems, and the hopping time optimization. Section~\ref{sec:res} describes our results and discussion, and Section~\ref{sec:con} summarizes our conclusions.

\section{Model and Methods}
\label{sec:methods}
\subsection{Model}
We study a system of two-dimensional repulsive discs confined to a narrow two-dimensional channel that extends longitudinally along the $x$-axis and has a radius, $R$, along the $y$-axis, where the origin of the coordinate system is located at the center of the channel. The particles interact through an inverse-power-law potential,
\begin{equation}
U(r_{ij})=
\epsilon\left(\frac{\sigma}{r_{ij}}\right)^\alpha, \quad 5 \le \alpha \le 100\mbox{,}
\label{eq:urij}
\end{equation}
where $r_{ij}=|r_i-r_j|$ is the distance between particles $i$ and $j$, $\alpha$ is the power-law exponent describing the repulsive potential, $\sigma$ is the particle diameter,  $\epsilon$ is the interaction strength, and the interaction is cut off at $r_c=2.5\sigma$.  We use the quantity $1/\alpha$ as a measure of particle softness because it tends to zero as $\alpha\rightarrow\infty$,  describing the hard sphere potential, and increases as the potential allows greater particle overlap. The particles interact with the walls of the channel as hard particles, so the particle-wall interaction is given by,
\begin{equation}
U_{W}({y}_{i})=
\begin{cases}
0 \ \ \textup{if} \ \lvert {y}_{i}\rvert \leq R-\sigma/2, \\
\infty \ \textup{if}  \ \lvert {y}_{i}\rvert > R-\sigma/2,
\end{cases}
\label{eq:HD_wall_interaction}
\end{equation}
where $y_{i}$ is the $y$-coordinate of particle $i$. Our simulations are carried out using reduced units. We also study the equivalent two-dimensional hard sphere system under the same conditions used to study the soft particles, so we can make a direct comparison.

\subsection{Transition State Theory for Hopping times} The particles in a fluid confined to a quasi-one-dimensional channel are arranged in a single file so that each particle is caged by its (first) nearest neighbour in each direction along the channel. In order to diffuse normally in the long time limit, the particles must hop past each other. This hopping is an activated process because the excluded volume interactions of the particles and the wall restrict the configuration space available to the particles as they pass, giving rise to an entropic free energy barrier. Transition state theory, which assumes that all the trajectories that cross through the transition state are reactive, i.e., recrossing events are ignored, provides an upper estimate for the transition rate in an activated process given by,~\cite{Bennet_Algorithms_1977,Chandler_Statistical_1978,Ruiz-Montero_Efficient_1997,ten_Wolde_Numerical_1999}
\begin{equation}
 k_{TST}=\frac{\left <| v_c^*|\right>_{eq}}{2}P_0(x_c^*)\mbox{,}\\
\label{eq:crossing_rate}
\end{equation}
where $P_0(x_c^*)$ is the probability density of the system located at the transition state, denoted $x_c^*$, and $v_c^*$ is the velocity of the system along the reaction coordinate as it passes through the transition state. The TST hopping time is then given by,
\begin{equation}
\tau_{hop}(TST)=\frac{1}{k_{TST}}\mbox{.}
\label{eq:t-hop_k-TST}
\end{equation}

The free energy barrier associated with a particle hopping past one of its neighbours to escape its cage is calculated using the small system isobaric-isothermal ($n,p,T$) ensemble developed by Corti et al.~\cite{Corti_Deriving_1998,Corti_Isothermal_2001} In general, the method involves considering the properties of a small system of $n$ particles confined to a small volume, $v$, immersed in a larger system of $N-n$ particles with volume $V-v$, both at a fixed temperature, $T$. One of the $n$ small system particles, denoted the shell particle, is constrained to a region $dv$ at the boundary of the small system, defining the small system volume. This avoids an over counting of configurations associated with the fluctuation of $v$ when integrating over the degrees of freedom of the large $N-n$ system to yield a small system constant pressure ensemble that does not rely on the use of a system size dependent volume scale.

\begin{figure}[ht]
\includegraphics[width=3.0in]{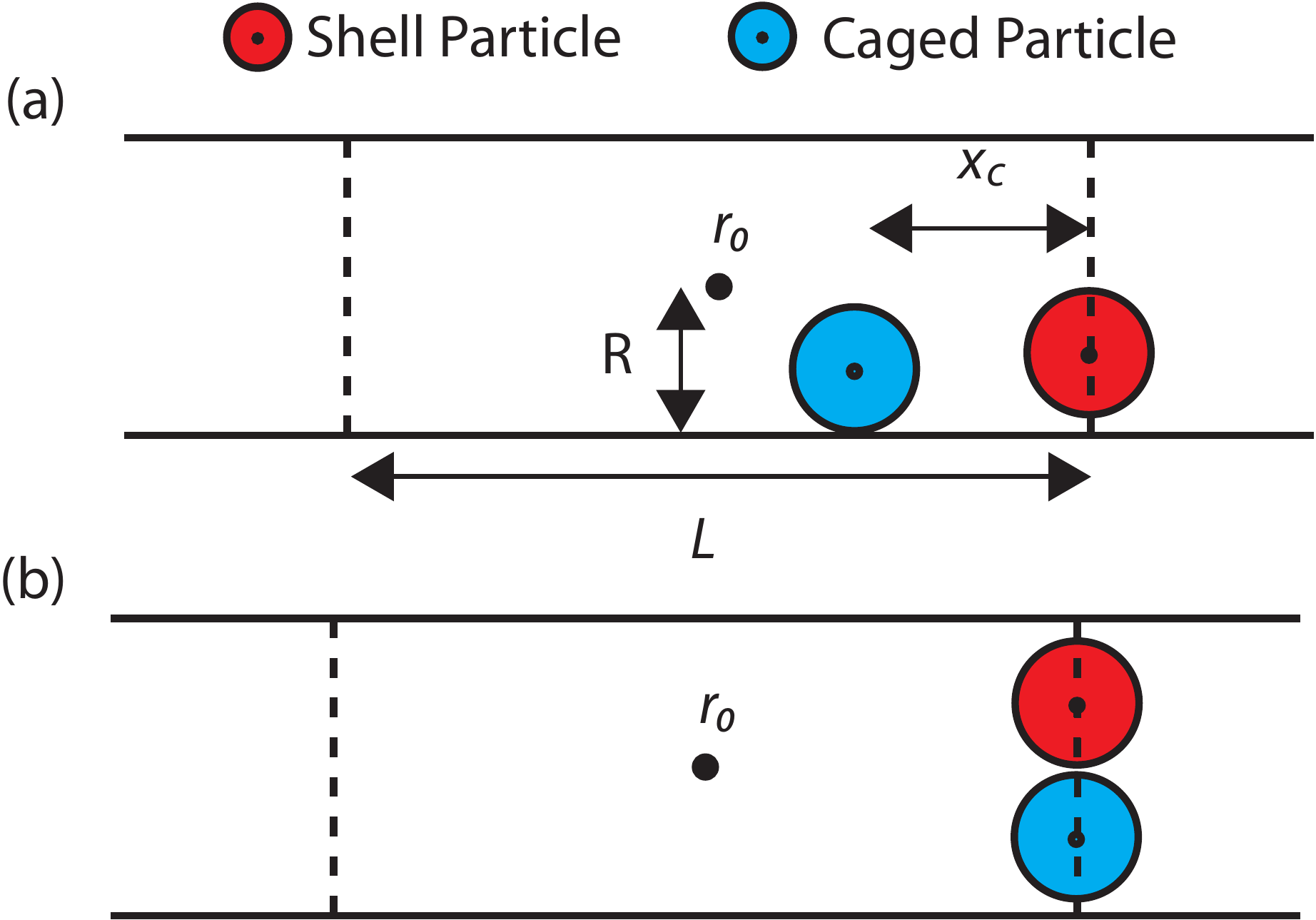}
\caption{(a) A schematic representation of the two-dimensional small $n,p_l,T$ system consisting of a cage with length $L$ and radius $R$, centered at the origin $r_0$. The shell particle is located at $L/2$ and defines the volume of the system. The reaction coordinate, $x_c$, is the longitudinal distance of the cage particle from the shell particle. (b) A configuration of the particles in the transition state where $x_c=0$.}
\label{fig:model}
\end{figure}

Here, we provide a brief description of the small system isobaric-isothermal ensemble in the context of its application to hopping times.~\cite{Ahmadi_Diffusion_2017} Figure~\ref{fig:model}(a) shows the system, consisting of $n=2$ particles, at a constant external longitudinal pressure, $p_l$, and fixed $T$. The small system has a length, $L$, and radius, $R$, with the center located at $r_0$. The shell particle (particle one) is confined to a region $2\,R\,dL$ at $x_1 = L/2 > 0$ to define the volume. The partition function for the small system given by,
\begin{equation}
\Delta= \int_L Q_{n,v,T}^{*} e^{-\beta p_l 2R L }\, 2\,R\, dL\mbox{,}
\label{eq:Delta2}
\end{equation}
where $Q^*_{n,v,T}$ is the canonical partition function of the small $n,v,T$ system with one shell particle and $\beta=1/k_BT$, where $k_B$ is the Boltzmann constant. In obtaining Eq.~(\ref{eq:Delta2}), it has been assumed that the interactions between the particles in the small and large systems are negligible.

Particle two in the system represents the caged particle, and we define a reaction coordinate for the hopping process as the axial separation between the particles, $x_c=x_1-x_2$, so that the transition state occurs when they are side by side in the channel, i.e., with $x_c=0$, as shown in Fig.~\ref{fig:model}(b). The reaction coordinate partition function is then given by,
\begin{equation}
\Delta_{x_{c}^{\prime}} \; dx_{c}= \int_{L} Q_{n,v,T}^{*} \; e^{-\beta p_l 2R L } \; \delta (x_{c}^{\prime}-x_{c}) 2R\;dLdx_c\mbox{,}\\
\label{eq:deltaxc0}
\end{equation}
where $\delta (\cdot)$ is the Dirac delta function. Integrating Eq.~\eqref{eq:deltaxc0} yields the partition function,
\begin{equation}
\Delta= \int_{x_{c}} \Delta_{x^{\prime}_{c}} \; dx_{c}\mbox{.}
\label{eq:deltaxc1}
\end{equation}
The probability of finding the caged particle at $x_{c}^{\prime}$ is then,
\begin{equation}
P_0(x_{c}^{\prime}) \; dx_{c}=\frac{\Delta_{x_{c}^{\prime}} \; dx_{c}}{\Delta}\mbox{,}\\
\label{eq:pxc}
\end{equation}
where the probability density, $P_0(x_{c}^{\prime})$, is normalized so that
\begin{equation}
\int_0^{\infty} P_0(x_{c}) \; dx_{c}=1\mbox{.}\\
\label{eq:intpxc}
\end{equation}
The Gibbs free energy barrier for hopping, which is defined as the work required to bring the caged particle from anywhere in the cage to the transition state is given by,
\begin{equation}
\beta\Delta G^*=-\ln P^*,
\label{eq:dg0}
\end{equation}
where
\begin{equation} 
P^*=\int_0^{x_c^{*}}P_0(x_c)dx_c\mbox{,}\\
\label{eq:pstar}
\end{equation}
and the range of the reaction coordinate, $x_c=0$ to $x_c^*$, is the transition state region. It should also be noted that we select $x_c^{*}$ to be small compared to the size of the motion required to take the particle from one cage to the other so that configurations that have little or no chance of crossing the barrier in the small time limit are excluded from the transition state.

The free energy barrier for hopping is obtained from a Monte Carlo (MC) simulation as follows: Particle one, the shell particle, is located at $L/2$, and particle two is placed within the cell between $-L/2$ and $L/2$. For each MC move, a particle is randomly selected and moved by $\delta x$ and $\delta y$, up to a maximum displacement of $|\Delta x|=0.06\sigma$ and $|\Delta y|=0.12\sigma$. The move is immediately rejected if the trial displacement causes the particle to overlap with the hard wall or if particle two is moved outside the cage. If particle one is moved, the cell length is  changed by  $2\delta x$ and the position of particle two is re-scaled to ensure it remains within the simulation cell. Because the position of the shell particle defines the system volume, the position of the shell molecule must be positive during the simulation to ensure $V>0$. The MC probability of accepting the trial move from the old to the new configurations is given by~\cite{Corti_Monte_2002,Ahmadi_Diffusion_2017}
\begin{equation}
\begin{split}
\mbox{acc}&\mbox{(old}\rightarrow\mbox{new)}=\mbox{min} (1,\exp \left\{\right. -\beta[U^{\mbox{\small{new}}}-U^{\mbox{\small{old}}}]\\
&-\beta p_l 2R[L^{\mbox{\small{new}}}-L^{\mbox{\small{old}}}]+(n-1)\ln[L^{\mbox{\small{new}}}/L^{\mbox{\small{old}}}] \left.\right\} ) \mbox{.}
\end{split}
\label{eq:mcacc_npt}
\end{equation}
For each system, the value of $\Delta G$ is obtained as an average over 20 independent runs. For each run, $2 \times 10^6$ MC cycles are used to reach equilibrium, and the free energy barrier is calculated over the next $8\times 10^8$ MC cycles, where an MC cycle consists of $n=2$ MC moves. We sampled configurations every 1000 MC cycles to ensure that they are not correlated. The probability is calculated by building a normalized histogram of all configurations along the reaction coordinate, where we have used bin sizes of $\Delta x_c=x_c^*=0.05\sigma$, and the free energy barrier is calculated using the probabilities. The error bars are calculated as the standard deviation of $\Delta G^*$ over the 20 runs. Our simulations consider systems with $\alpha$ in the range 5--100 and $R/\sigma$ in the range 1.01--1.10 under conditions with $\beta p_l=1$ and $\beta=1$. The probability density at the transition state, $P_0(x_c^*)$, is used in Eq.~\eqref{eq:crossing_rate} to calculate the transition rate, and the hopping time is obtained by dividing the transition state probability, $P^*$, by $\Delta x_c$.

The average velocity at which the system crosses the barrier is defined by
\begin{equation}
\left<| v_c^*|\right>=\left<\frac{|x_{c}(t+\Delta t)-x_{c}(t)|}{\Delta t}\right>\mbox{,}
\label{eq:ave_flux}
\end{equation}
where $x_c(t+\Delta t)$ and $x_c(t)$ are the values of the reaction coordinate close to the transition state. To obtain $\left<|v_c^*|\right>$, we performed simulations in the canonical ($n,V,T$) ensemble using the standard Metropolis MC scheme~\cite{Frenkel_Understanding_2002} to move the particles as a simple approximation to Brownian motion.~\cite{Patti_Brownian_2012} Each MC move involves randomly selecting one of the two particles and moving it randomly by a step $\delta x$ and $\delta y$ up to a maximum displacement of $|\Delta x|=0.06\sigma$ and $|\Delta y|=0.12\sigma$. The move was rejected if the trial displacement caused any of two particles to overlap with the hard wall; otherwise the move was accepted according to the standard Metropolis MC probabilities. A Monte Carlo cycle involves $n=2$ attempted MC moves and defines the unit of time. We then used $\Delta t=1$ in Eq.~\eqref{eq:ave_flux}. The starting configuration for each run, with $x_c(t=0)<0.05$, was taken from the free energy simulations described in the previous section, and average quantities were calculated over 1000 distinct initial configurations.

\subsection{Hopping Time Measurement}
In order to test the predictions of our TST, we directly measure the hopping time in a series of canonical ($N,V,T$) simulations consisting of $N=500$ particles confined to a channel of radius, $R$, and length, $L$. Periodic boundary conditions are used in the longitudinal direction, and the particle dynamics is the same as that described in the previous section. For each MC move, a particle is randomly selected and moved in a random direction by a step $\delta x$ and $\delta y$ up to a maximum displacement of $|\Delta x|=0.06\sigma$ and $|\Delta y|=0.12\sigma$. An MC cycle involves $N$ MC move attempts and defines the unit of time, $t$, in the simulation. Particles are initially placed uniformly along the channel but randomly across the width of the channel. Then $3\times 10^7$ MC cycles are performed to equilibrate the system before collecting data over the next $5 \times10^7\; \textup{to}\; 3\times10^8$ MC cycles in order to be long enough for the average hopping time to converge.  After equilibrium is achieved, the cage for each particle is defined by their immediate right and left neighbours, and the initial hopping time is set to zero for all the particles. The number of MC cycles that it takes for each particle to escape their cage is counted as their hopping time. After each hopping event, the hopping time for the particle is reset to zero and the new cages are identified. The hopping time, $\tau_{hop}$, is calculated as an average over all hopping times recorded for all particles. We also performed simulations with $N=100-800$ to check for system size effects on the hopping time calculations, and these additional simulations confirm $N=500$ is large enough to account for system size effects.

\begin{figure}[ht]
\includegraphics[width=3.0in]{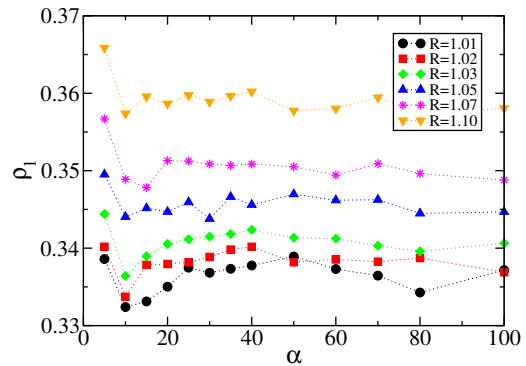}
\caption{Linear density, $\rho_l$, obtained from $N,p_l,T$ simulations as a function of $\alpha$ for different channel radii.}
\label{fig:density}
\end{figure}

The appropriate linear density, $\rho_l=N/L$ for each system ($R,\alpha$) studied is determined by performing an $N,p_l,T$ simulation, where $N=500$, $p_l/k_BT=1$. In the limit of large system sizes, the standard constant pressure simulation method and the shell particle method yield the same results,~\cite{Corti_Monte_2002} so we continued to use the shell particle method.  An MC trial move follows that outlined for the free energy calculations (Eq.~\eqref{eq:mcacc_npt}), with the maximum step size of $|\Delta x|=0.06\sigma$ and $|\Delta y|=0.12\sigma$, and results in an MC acceptance ratio of $85-92 \%$. The average $\rho_l$ for the system  is measured over $5 \times 10^7$ MC cycles, after the system reaches equilibrium over $10^7$ MC cycles, and data are sampled every $10\,000$ MC moves to ensure that they are not correlated. Figure~\ref{fig:density} shows that the densities increase as the channels expand but remain relatively constant as a function of $\alpha$.

\section{Results and Discussion}
\label{sec:res}
\subsection{Hopping times}

\begin{figure}[htbp]
\includegraphics[width=3.0in]{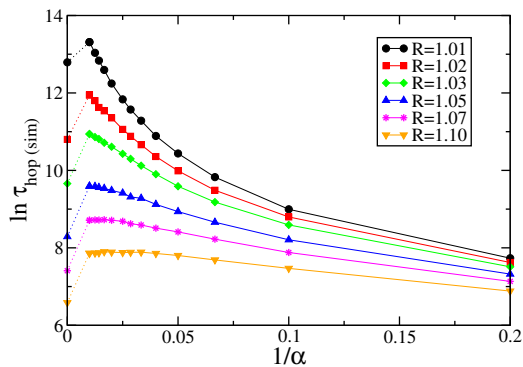}
\caption{$\ln \tau_{hop}$ for the hopping times measured from simulation as a function of $1/\alpha$ for different channel radii.}
\label{fig:mtimes}
\end{figure}

The hopping times obtained directly from the large system simulations provide a measure by which we can qualitatively examine the influence of particle softness on diffusion. Figure~\ref{fig:mtimes} shows that, as soft particles become harder (decreasing $1/\alpha$), $\tau_{hop}$ initially increases, doing so more rapidly for the narrower channels. This is consistent with our expectation that particle hopping in a channel should be more difficult for harder particles. We also know that the hopping time must diverge in the limits $R\rightarrow\sigma$ and $1/\alpha\rightarrow 0$ because perfectly hard particles are unable to pass each other when the channel radius decreases below the passing threshold, causing the fluid to undergo a dynamic transition from normal diffusion to SFD. However, we also find that the hopping times exhibit an unexpected maximum, suggesting that there are cases where the softer particles diffuse more slowly than the harder particles. The maximum, which is directly observable for the wider channels, sharpens and moves to lower values of $1/\alpha$ as the channels narrow until it is no longer visible within the range of  $1/\alpha$ studied. In these narrow channel cases, the hopping times grow larger than those measured for the perfect hard discs model, suggesting that the maximum may still occur at lower values of $1/\alpha$ than were studied.

\begin{figure}[htbp]
\includegraphics[width=3.0in]{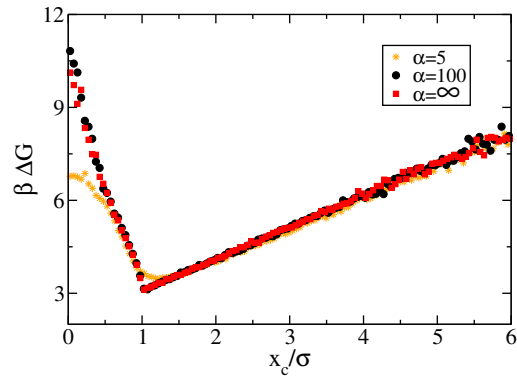}
\caption{Gibbs free energy as a function of the reaction coordinate, $x_c$, for systems with $\alpha = 5$ (stars), $\alpha=100$ (circles), and $\alpha=\infty$ (squares) with $R_p/\sigma=1.01$.}
\label{fig:rc}
\end{figure}

We now examine the properties of the free energy barrier to hopping. Figure~\ref{fig:rc} shows the Gibbs free energy as a function of separation of the two particles along the reaction coordinate for systems with different values of $\alpha$. At large $x_c$, the interaction between the particles is negligible, and they are both able to move freely across the entire diameter of the channel. However, constraining the particles to a particular $x_c$ puts a lower limit on the fluctuations available to the cell in the $n,p_l,T$ ensemble. The linear increase in the free energy with increasing $x_c$ is related to the work performed against the external pressure as the particles are separated, forcing the cage to increase in volume. The same effect is observed for two ideal gas particles.~\cite{Ahmadi_Diffusion_2017} The free energy goes through a minimum at $x_c\approx\sigma$ and begins to increase as the repulsive interaction between the particles begins to restrict their available configuration space. The minimum is sharp for the harder particles, but with decreasing $\alpha$, the softer, longer range interactions lead to a shallower and softer minimum as well as a slower increase toward the maximum near the transition state at $x_c=0$. When the channels are very narrow, the maximum is located very close to the geometrically defined transition state, but this is not always the case.~\cite{Ahmadi_Diffusion_2017} The particles studied here are circular, so the greatest degree of configurational restriction occurs when $x_c=0$, but as the channel becomes wider, the influence of the pressure-volume work become significant, and this can lead to the appearance of the maximum before the geometric transition state is reached. Nevertheless, it is important to recognize that it is the probability of being at the geometric transition state that contributes to the hopping time because the particles are unable to diffuse in the long time limit unless they exchange positions along the channel.

\begin{figure}[htbp]
\includegraphics[width=3.0in]{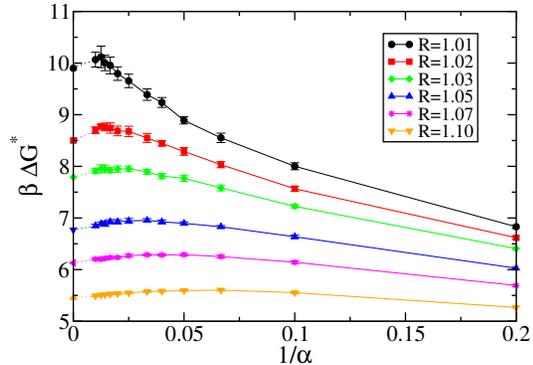}
\caption{Gibbs free energy barrier, $\beta\Delta G^*$, as a function of $1/\alpha$ for different channel radii.}
\label{fig:barrier}
\end{figure}

\begin{figure}[ht]
\includegraphics[width=3.0in]{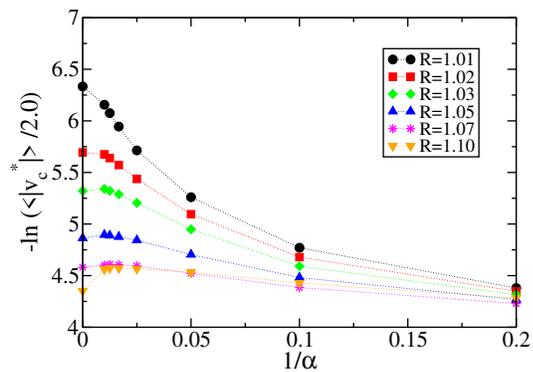}
\caption{The prefactor term, $-\ln(\left<| v_c^*|\right>/2.0)$, as a function of $1/\alpha$ for different channel radii.}
\label{fig:prefactor}
\end{figure}

Figure~\ref{fig:barrier} shows the height of the free energy barrier for hopping exhibits the same general trends as $\ln\tau_{hop}(sim)$ measured from the simulation, and moreover we are able to directly identify the location of the free energy maximum for all channel radii. Measurements of the prefactor term (see Fig.~\ref{fig:prefactor}) also account for the disappearance of the maxima in the hopping times for narrow channels. For wider channels, the prefactor is relatively independent of the interaction potential, so the hopping times essentially follow the behaviour of the free energy. However, for narrower channels, the prefactor term becomes strongly potential dependent, and its increase outweighs the effect of the decreasing free energy. This highlights the importance of understanding the properties of the prefactor even in a regime where the free energy barriers are high and are expected to dominate the overall behaviour. Figure~\ref{fig:prefactor} also shows that the prefactor is relatively independent of channel width for the softer particles but is highly channel width dependent for the harder particles.


\begin{figure}[htbp]
\includegraphics[width=3.0in]{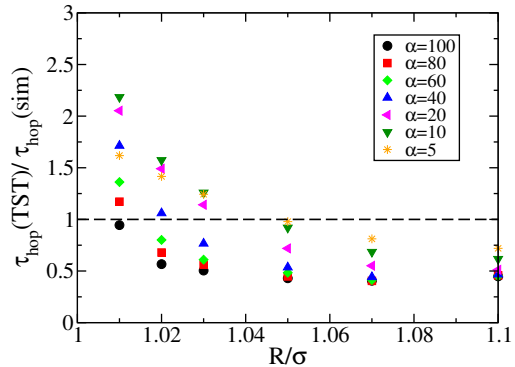}
\caption{Comparison of predicted hopping time, $\tau_{hop}(TST)$ with the measured hopping time, $\tau_{hop}(sim)$, for different values of $\alpha$ and $R$. The dashed line represents perfect agreement. }
\label{fig:tvst}
\end{figure}

Figure~\ref{fig:tvst} shows that our TST approach generally predicts the hopping times $\tau_{hop}(TST)$ to within a factor of two, underestimating the times for wide channels and overestimating them for narrow channels. Transition state theory should underestimate the hopping time because it assumes that all trajectories that cross through the transition state lead to a hopping event. Improvements that account for trajectory recrossing, where the particle re-enters its original cage, correct terms associated with the kinetic prefactor and lead to longer hopping times. However, our two particle approach makes some additional assumptions that may lead to further variations from the measured value. The construction of the two particle model, through the mapping of the large canonical ensemble onto the small system $n,p,T$ ensemble is formally exact, but when we perform our calculations, we neglect the interactions between the system and its surroundings, leading to errors in the free energy calculations. In particular, we expect the free energy calculations to be less accurate for the systems with softer, longer range potentials because we neglect the interactions outside of the cell. As the channels become narrower, the fluid structure becomes more single file, reducing the role of long range interactions and improving the accuracy of the free energy calculations. It is important to note that the prefactor calculations are also performed using just two particles; this ignores the role second neighbours may play in blocking the escape of a particle from its cage. Nevertheless, the simplicity of our two particle analysis combined with its ability to capture the general trends in the hopping times for the particles suggests that it could be useful in determining the factors that influence diffusion in highly confined fluids.

\subsection{Transition State Partition Function}
\begin{figure}[htbp]
\includegraphics[width=3.1in]{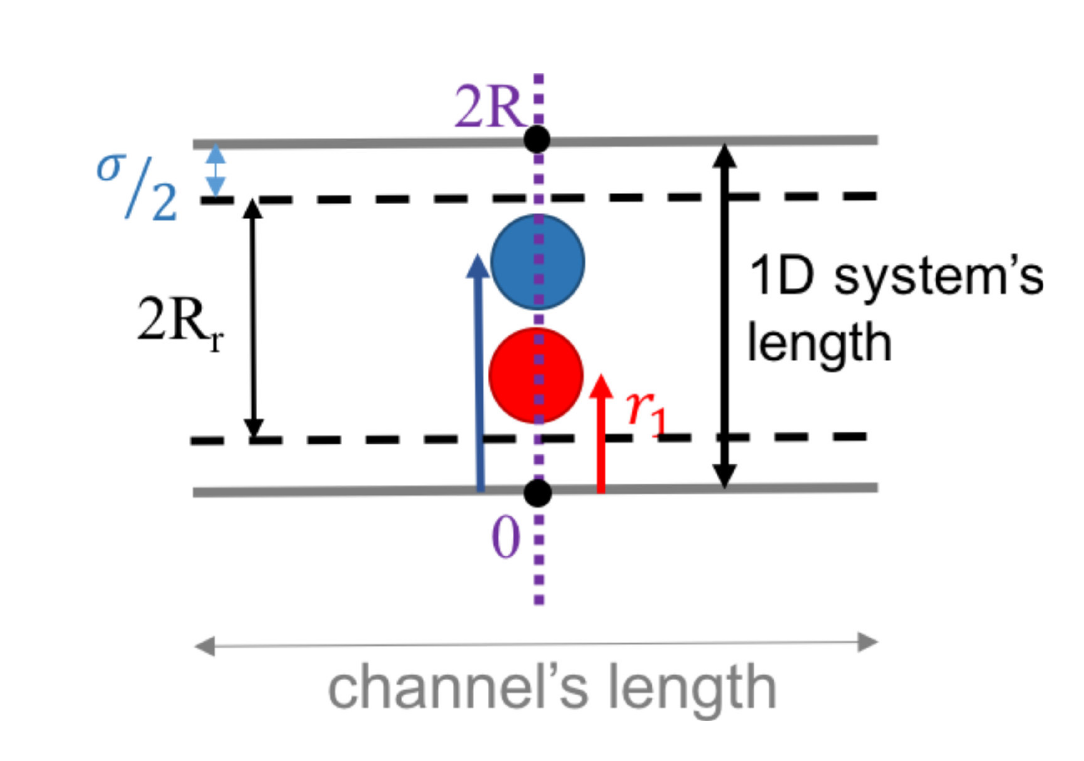}
\caption{A schematic representation of the one-dimensional transition state.}
\label{fig:1dtrans}
\end{figure}

To gain insight into the origin of the maximum in the barrier height as a function of particle softness, we study the ensemble of states associated with the transition state, which for the current system is described by the partition function for two particles confined to a one-dimensional line (see Fig.~\ref{fig:1dtrans}). The one-dimensional transition state partition function for the system can be written as,
\begin{equation}
Q_{1d}=\frac{2}{2!\Lambda^2}\int_0^{2R_r}dr_1\int_0^{2R_r-r_1}e^{-\beta U(r_{ij})}dr_{ij}\mbox{,}\\
\label{eq:qtst}
\end{equation}
where $\Lambda=(2mk_BT/h^2)^{1/2}$ is the thermal de Broglie wavelength for a particle with mass $m$, $h$ is Planck's constant, $R_r=R-\sigma/2$ is the reduced channel radius accessible to the centers of the particles due to the hard wall interaction, $r_{ij}=r_2-r_1 > 0$ is the particle separation, $U(r_{ij})$ is the interaction potential given by Eq.~\eqref{eq:urij}, and the factor of two accounts for the need to consider both possible particle orders on the line. The Helmholtz free energy for the transition state, relative to two ideal gas particles, is then given by,
\begin{equation} 
\beta\Delta F^*_{1d}=-\ln\left(Q_{1d}/Q_{ig}\right)\mbox{,}\\
\label{eq:dftst}
\end{equation}
where $Q_{ig}=(1/2!\Lambda^2)(2R)^2$ is the ideal gas partition function, noting that the ideal gas particles have no diameter and hence are able to sample the entire width of the channel. The energetic and entropic contributions to the free energy, $F=E-TS$,  can then be obtained using the usual canonical ensemble expressions, $\left<E\right>=k_BT^2(\partial \ln Q/\partial T)_{N,V}$ and $S=k_B\ln Q+\left<E\right>/T$, respectively.

\begin{figure}[t]
\includegraphics[width=3.0in]{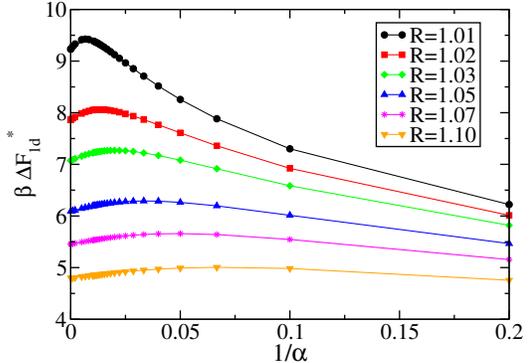}
\caption{Helmholtz free energy for the one-dimensional transition state ensemble as a function of $1/\alpha$ for different channel radii.}
\label{fig:1dfe}
\end{figure}

Figure~\ref{fig:1dfe} shows that, despite its simplicity, the one-dimensional transition state partition function captures the key features of the effect of the interaction potential on the hopping free energy barrier, including the presence of the maximum and how it evolves as a function of channel radius. However, it is important to recognize that the transition state partition function cannot be used directly in the TST expression to yield the hopping time because it does not give the probability of finding the system at the top of the barrier, but it clearly contains the relevant details on how the properties of the transition state vary as the system parameters vary. 

\begin{figure}[t]
\includegraphics[width=3.0in]{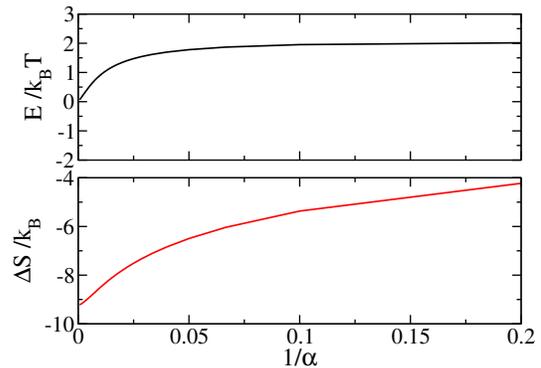}
\caption{The energy, $E/k_BT$, and entropy relative to an ideal gas, $\Delta S/k_B$, of the one-dimensional transition state ensemble as a function of particle softness for a channel with $R=1.01$.}
\label{fig:1dentropy}
\end{figure}

The model also reveals that the maximum in the height of the hopping barrier as a function of particle softness results from a competition between the energy and entropy in the transition state as the interaction potential changes (see Fig.~\ref{fig:1dentropy}). The barrier for the hard disc system is entirely entropic in nature, with the excluded volume interaction between the particles and the wall restricting the accessible configurations space for the two particles. As the particles become softer, both the energy and the entropy relative to the ideal gas begin to increase because the particles can effectively overlap, but for small $1/\alpha$, the energetic cost increases more rapidly than the entropy, causing an overall  increase in the height of the barrier. For larger values of $1/\alpha$ the situation is reversed. The energy essentially plateaus, and the entropy increasingly dominates, leading to an overall decrease in the hopping free energy barrier as the particles continue to become softer.

\subsection{Optimization}
Using the hopping time approach to study diffusion in quasi-one-dimensional systems is attractive because all the molecular details that influence mobility are contained within a single parameter, $\tau_{hop}$, suggesting it could be a useful target for the optimization of diffusion in these systems. Our TST then provides a simple and efficient way to evaluate the $\tau_{hop}$ as part of an optimization scheme.

To demonstrate the general principle, we use the pythOPT global optimization problem-solving software environment~\cite{Voss:Thesis:pythOPT} to search for the value of the particle softness $(1/\alpha)$ that maximizes the hopping barrier for a given channel radius. This process corresponds to finding the maxima in Fig.~\ref{fig:barrier} and approximately identifies the conditions under which diffusion is slowest. The pythOPT environment offers multiple global optimization solvers as well as a suite of benchmark problems and routines for performance analysis. The solver used in this study was the Guaranteed Convergence Particle Swarm Optimization (GCPSO).~\cite{vandenBergh2002} At each step in the search, the MC free energy calculation outlined in Section~\ref{sec:methods} is used to obtain the figure of merit. This obviously ignores the effects associated with the prefactor, but they can be included in the current analysis in a straightforward way. CGPSO simulations used 50 swarm particles and one million function evaluations. Each DG simulation took approximately five minutes to execute, and 40 simulations were used for each function evaluation. The optimization for each radius required approximately five CPU-days.

To compare the optimization results with estimates obtained from our systematic set of simulations, we fit a quadratic polynomial to the free energy as a function of $1/\alpha$ to the data in Fig.~\ref{fig:barrier} for each channel radius for the five closest points the free energy maximum. Table~\ref{tab:DG_maxima_table} shows that there is excellent agreement between the maxima obtained from pythOPT and the systematic study for $\beta\Delta G^*$. As can be expected from the sensitivity of optimization problems~\cite{Heath2018}, the agreement is not as good for the location $1/\alpha$ of the maxima, but the values are still close given the large degree of fluctuation in our free energy estimates. For example, for $R=1.01$, the fluctuation in $\beta\Delta G^*$ is on the order of $0.20\ kT$, and this could include values of $1/\alpha$ in the range $0-0.17$.

\begin{widetext}
\begin{table}[htbp]
\centering
\begin{tabular}{p{2.5cm}p{2.5cm}p{2.5cm}p{2.5cm}p{2.5cm}}
\hline
& \multicolumn{2}{l}{Global Optimization} & \multicolumn{2}{l}{Quadratic Fit}\\
\hline
 $R$ & $1/\alpha$ & $\beta\Delta G^*$  & $1/\alpha$ & $\beta\Delta G^*$   \\
\hline\hline
 1.01  & 0.0107 & 10.095 & 0.0104 & 10.084\\
 1.03  & 0.0123 & 7.978 & 0.0165 & 7.947\\
 1.05  & 0.0215 & 6.957 & 0.0292 & 6.950 \\
 1.07  & 0.0429 & 6.294 & 0.0425 & 6.289 \\
 1.10  & 0.0585 & 5.603 & 0.0626 & 5.601 \\
\hline
\end{tabular}
\caption{Values of $1/\alpha$ and $\beta\Delta G^*$ at the maximum in the hopping free energy barrier.}
\label{tab:DG_maxima_table}
\end{table}
\end{widetext}

The example given here is relatively straightforward because we only vary one parameter and there is a well defined maximum in the free energy surface. However, it demonstrates the principle of using the TST approach, either by simply calculating the free energy or the full hopping time through the added calculation of the prefactor term, in a search for an optimal condition to control diffusion.  In general, an optimization for the purposes of engineering design would involve the variation of a variety of parameters relating to channel diameter and particle--particle and particle--wall interactions.

\section{Conclusions}
\label{sec:con}
The hopping time, $\tau_{hop}$, is a phenomenological parameter that connects a local parameter, i.e., the time it takes for a particle to escape the cage of its neighbours, to the long time diffusion coefficient of the system. It contains information about the roles system parameters, such as the density and particle-particle and particle-wall interactions, play in the dynamics of quasi-one-dimensional fluids. We have shown that a transition state theory, where the free energy barrier for two particles attempting to pass is calculated in the small system isobaric-isothermal ensemble, provides quantitative predictions for $\tau_{hop}$ and is able to reveal interesting details regarding effect of inter-particle interaction in the dynamics of the system. This suggests that the hopping time TST approach could be an effective tool for the optimization and control of diffusion in nano- and micro-fluidic devices.

\acknowledgments
We would like to thank Natural Sciences and Engineering Research Council of Canada (NSERC) for financial support. Computing resources and support were provided by Compute Canada and WestGrid.


%

\end{document}